\documentclass[review]{elsarticle}

\usepackage{lineno,hyperref}
\modulolinenumbers[5]
\usepackage{algorithm}
\usepackage{amsmath}  
\usepackage{algorithmic}
\usepackage{amsfonts}
\usepackage{subfigure}
\usepackage{bm}
\usepackage{dsfont}
\usepackage{booktabs}

\makeatletter
\def\ps@pprintTitle{%
  \let\@oddhead\@empty
  \let\@evenhead\@empty
  \def\@oddfoot{\reset@font\hfil\thepage\hfil}
  \let\@evenfoot\@oddfoot
}
\makeatother

\begin{document}

\begin{frontmatter}

\title{MELAGE: A purely python based Neuroimaging software (Neonatal)}


\author[inibica]{Bahram Jafrasteh\corref{mycorrespondingauthor}}
\ead{jafrasteh.bahram@inibica.es}

\author[inibica,hpum]{Sim\'on Pedro Lubi\'an-L\'opez}
\ead{simonp.lubian.sspa@juntadeandalucia.es}
\author[inibica,hpum,uca]{Isabel Benavente-Fern\'andez}
\ead{isabel.benavente@uca.es}

\cortext[mycorrespondingauthor]{Corresponding author}

\address[inibica]{Biomedical Research and Innovation Institute of C\'adiz (INiBICA) Research Unit, Puerta del Mar University, C\'adiz, Spain}
\address[hpum]{Division of Neonatology, Department of Paediatrics, Puerta del Mar University Hospital, C\'adiz, Spain}
\address[uca]{Area of Paediatrics, Department of Child and Mother Health and Radiology, Medical School, University of C\'adiz, C\'adiz, Spain}

\begin{abstract}
MELAGE, a pioneering Python-based neuroimaging software, emerges as a versatile tool for the visualization, processing, and analysis of medical images. Initially conceived to address the unique challenges of processing 3D ultrasound and MRI brain images during the neonatal period, MELAGE exhibits remarkable adaptability, extending its utility to the domain of adult human brain imaging. At its core, MELAGE features a semi-automatic brain extraction tool empowered by a deep learning module, ensuring precise and efficient brain structure extraction from MRI and 3D Ultrasound data. Moreover, MELAGE offers a comprehensive suite of features, encompassing dynamic 3D visualization, accurate measurements, and interactive image segmentation. This transformative software holds immense promise for researchers and clinicians, offering streamlined image analysis, seamless integration with deep learning algorithms, and broad applicability in the realm of medical imaging.
\end{abstract}


\end{frontmatter}


\section{Introduction}\label{Intro}
Every year, approximately 15 million infants are born prematurely (before 37 weeks of gestation) worldwide \cite{Who}. In Europe alone, this accounts for around 500,000 infants annually, constituting approximately 10\% of all newborns \cite{haumont2012situation}. 
The foremost challenge in caring for very low birth weight premature infants today is to optimize their neurodevelopmental outcomes while reducing the risk of future disabilities. To achieve this, early detection of brain injuries is crucial. This emphasis on early diagnosis is paramount, given that clinical examinations are complicated by the underdeveloped neurological function in neonates \cite{longo2021neurodevelopmental}.

The development of imaging biomarkers during the neonatal period can address this pressing need for early diagnosis. Traditional clinical assessments are hindered by the limited neurological function in newborns. These biomarkers hold the potential to expedite the creation of neuroprotective strategies and the implementation of evidence-based care programs tailored to enhance brain development and promote neuroplasticity.

Neuroimaging methods, notably magnetic resonance imaging (MRI) and three-dimensional ultrasonography (US), prove to be powerful diagnostic instruments in neonatal medicine. These techniques not only aid in the identification of brain injuries but also provide the means to explore diverse parameters that hold the potential to function as predictive markers for the outcomes of preterm infants. Furthermore, MRI imaging, including T1-weighted and T2-weighted imaging, finds utility in the analysis of the adult human brain as well.

These methods enable the examination of various regions of interest (ROI) within the brain and the extraction of both simple and intricate parameters, including geometric characteristics. Two primary approaches are commonly employed:
\begin{itemize}
\item Automated image segmentation methods, often utilizing deep learning or artificial intelligence techniques \cite{soomro2022image, latif2021recent}. These methods offer automation and speed, but they may struggle to generalize effectively when applied to previously unseen images. Consequently, they typically require validation by domain experts.

\item Manual image segmentation, which entails a skilled expert segmenting images slice by slice. Several general-purpose software tools are available for manual image segmentation, such as 3D Slicer \cite{fedorov20123d} and ITK-SNAP \cite{yushkevich2006user}.
\end{itemize}
Fully manual image segmentation is a time-intensive process that demands the expertise of medical professionals. While it is widely regarded as the gold standard, some studies have reported that automatic segmentation methods can achieve comparable or superior accuracy compared to manual segmentation by an expert \cite{yepes2023eliminating}. However, it's worth noting that automatic segmentation often necessitates oversight and adjustments by an expert, emphasizing the importance of close collaboration between engineers and medical professionals to advance these methods.

Designing a software interface that balances user-friendliness for routine clinical use with advanced features for both manual and automatic segmentation remains a significant challenge. Most clinicians are hesitant to adopt non-intuitive and complex software solutions.

To address these challenges, we introduce MELAGE, a novel software platform for semi-manual segmentation, visualization, and correction. MELAGE is built entirely in Python, offering seamless integration with other Python libraries to facilitate automated segmentation while allowing for subsequent manual adjustments.

In this paper, we provide a concise overview of the core functionality of the MELAGE image analysis tool. The paper is structured as follows:

In Section \ref{features}, we highlight the capability of MELAGE which provides a brief discussion of the tool's key features and also advanced tools offered within MELAGE.
Finally, in Section \ref{conclusion}, we offer concluding remarks on MELAGE as an user-friendly and innovative neuroimaging tool for human brain analysis using MRI and Ultrasound imaging.

\section {Main features of MELAGE} \label{features}
\subsection{Basic}
MELAGE has several options for image analysis.
\subsubsection{Image enhancement}
At times, the presence of low-quality images, coupled with interobserver variability during manual segmentation, poses a significant challenge in achieving a dependable segmentation outcome. To mitigate this issue, we have incorporated image enhancement functionalities within our system. These capabilities encompass adjustments such as brightness and contrast modifications, the application of band-pass filters in the frequency domain, as well as Sobel edge detection \cite{kanopoulos1988design} and Hamming filters \cite{lowe1997spatially}. These enhancements collectively contribute to the robustness and accuracy of the segmentation process using MELAGE.
\begin{figure}[H]
\centering
\subfigure[]{
\label{fig:seg_a}\includegraphics[width=1\linewidth]{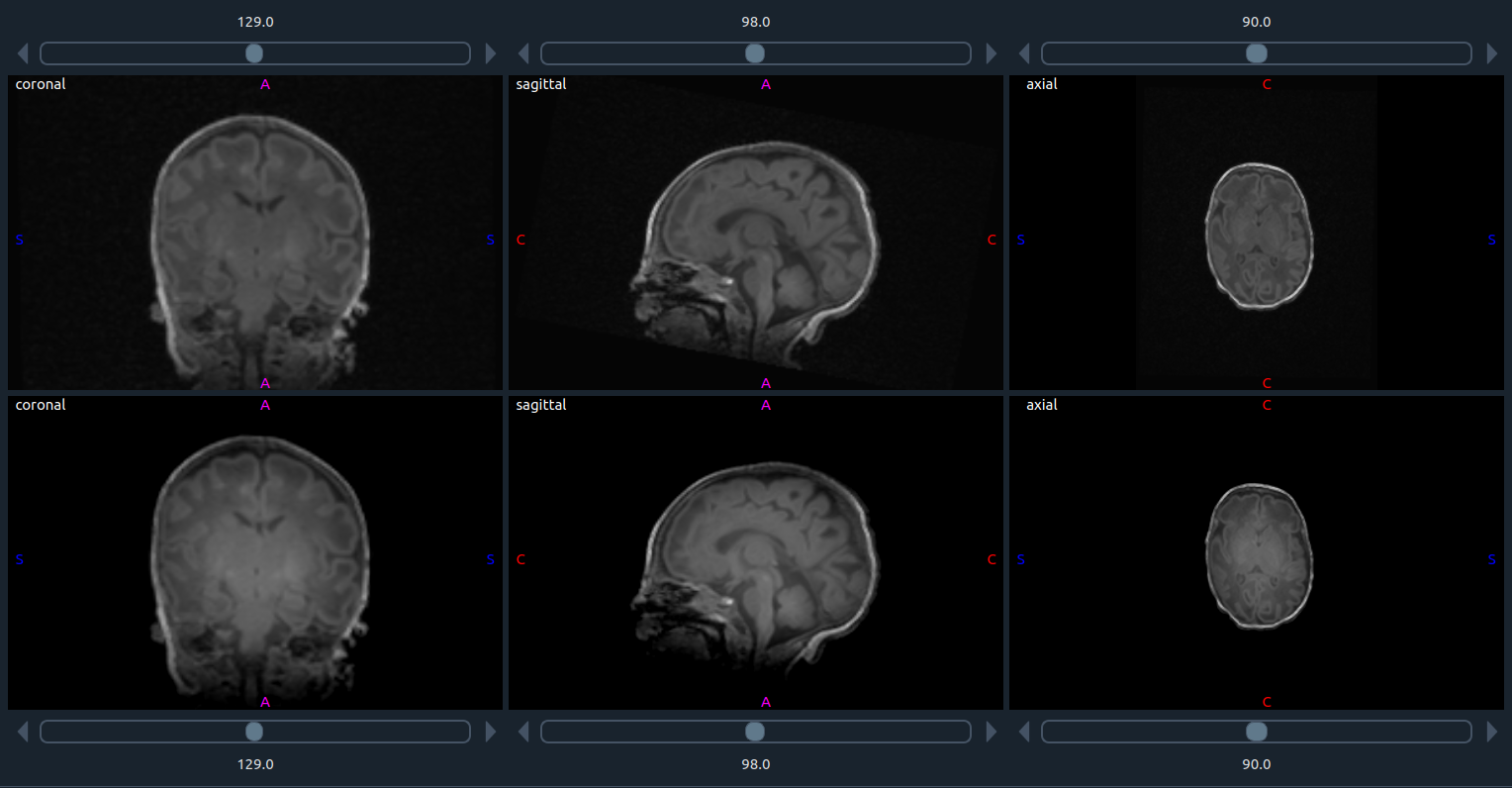}}
\caption{MRI image from a preterm patient at PMA 34.5 weeks. Top image: Original image; coronal, sagittal and axial planes are seen from left to right, Bottom Image: enhanced image by changing the image brightness and its contrast.}
\end{figure}
We have incorporated a widget that allows users to adjust brightness and contrast parameters. These adjustments take effect immediately, affording users the flexibility to fine-tune parameters to their specific requirements.

\begin{figure}[H]\label{fig_imageEn}
\centering
\subfigure[Original Slice from a brain Ultrasound image]{
\label{fig:seg_a}\includegraphics[width=0.4\linewidth]{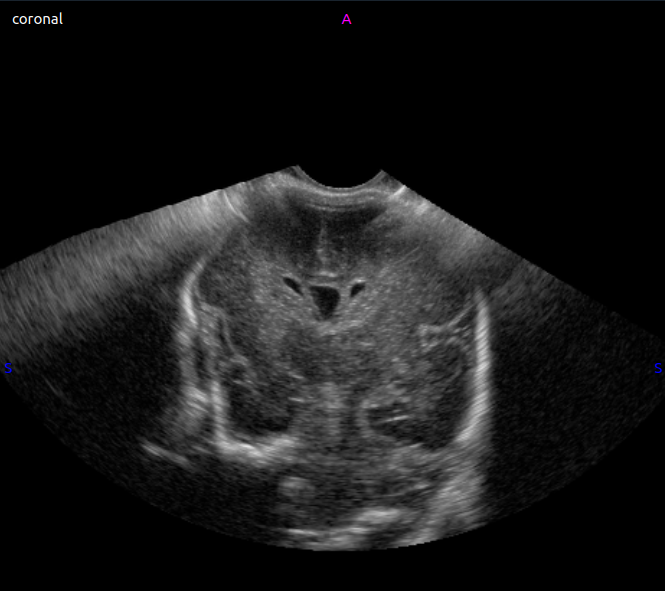}}
\subfigure[Brightness increase]{
\label{fig:seg_b}\includegraphics[width=0.4\linewidth]{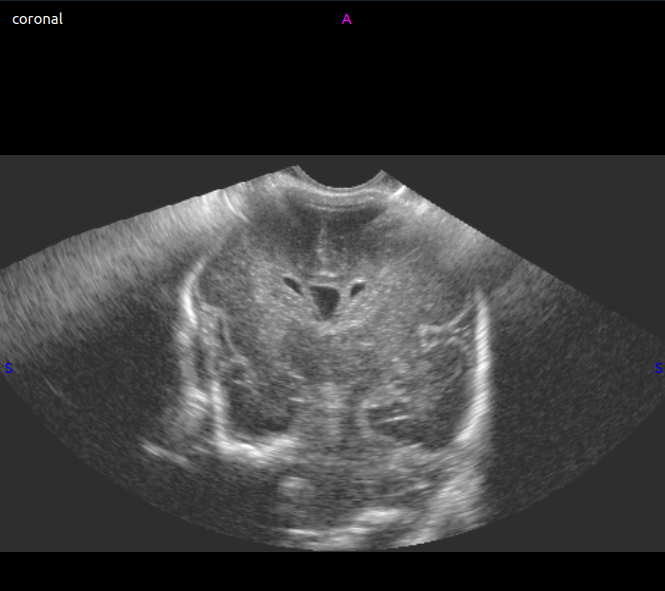}}\\
\subfigure[band pass filter]{
\label{fig:seg_b}\includegraphics[width=0.4\linewidth]{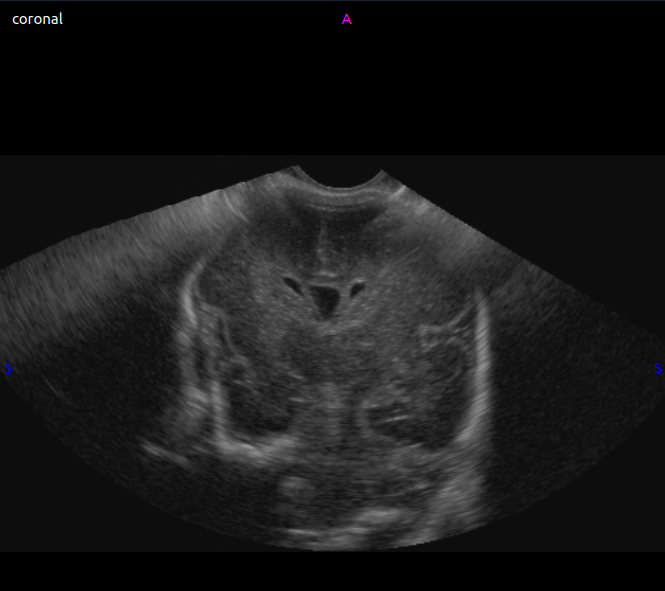}}
\subfigure[band pass filter with changing the image brightness]{
\label{fig:seg_b}\includegraphics[width=0.4\linewidth]{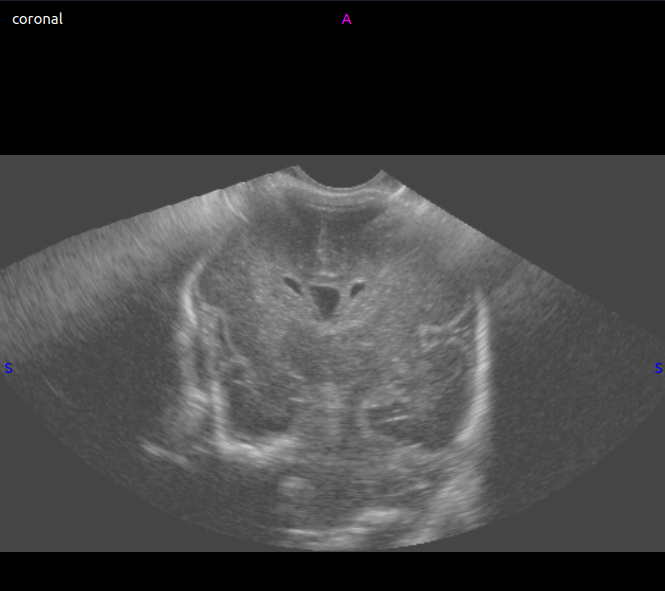}}
\caption{Image enhancement in a coronal slice taken from 3D image of a preterm neonate at HUPM.}
\end{figure}

Once the parameters are set, they are consistently applied across all slices and planes for the currently displayed image. In Figure \ref{fig_imageEn}, we illustrate image enhancement techniques applied to a slice taken from an ultrasound image.

\subsubsection{Mutual view}
\begin{figure}[H]
\centering
\subfigure[]{
\label{fig:mutual_view}\includegraphics[width=1\linewidth]{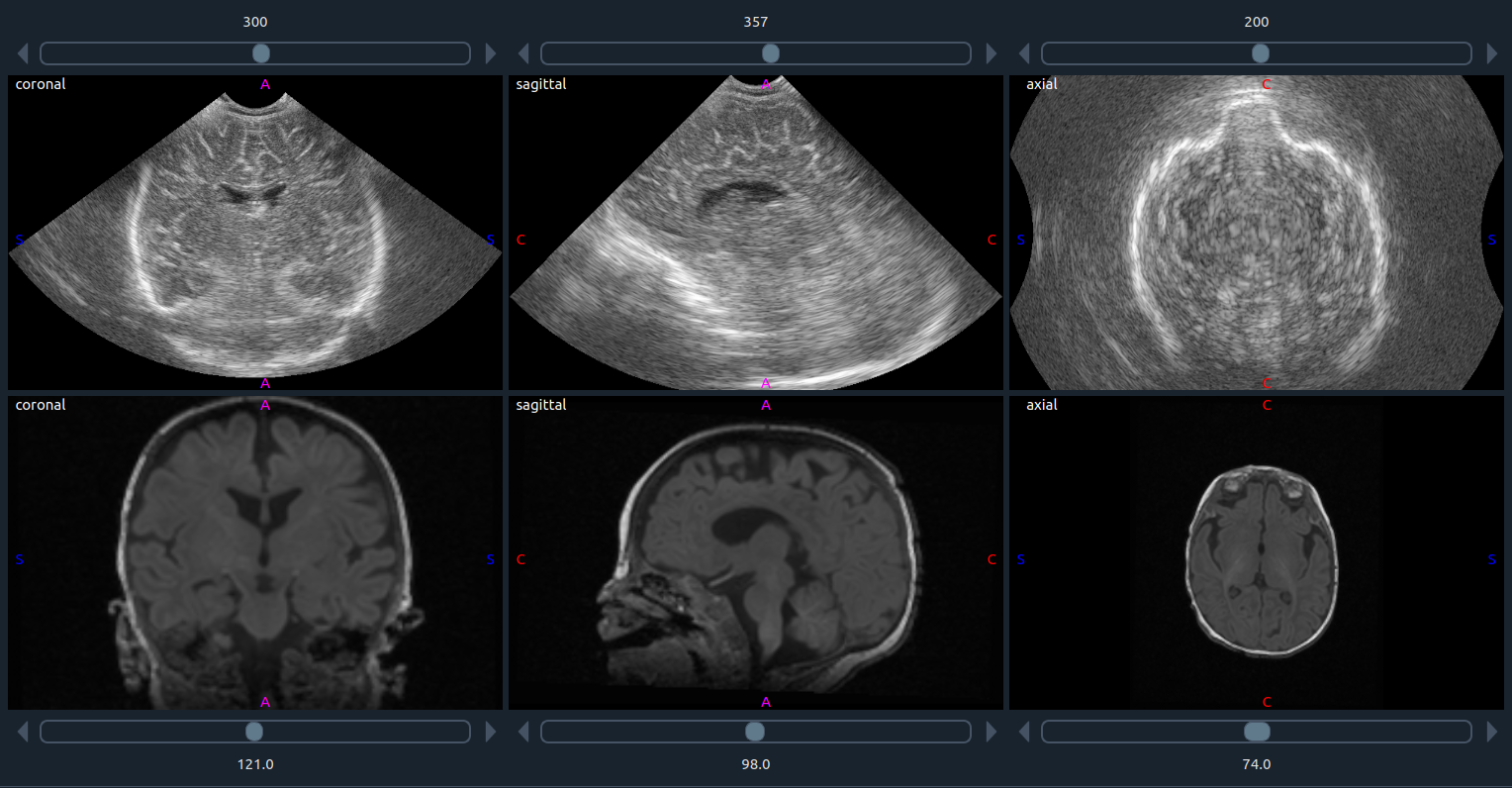}}
\caption{Top Image; Ultrasound image from a patient taken at PMA of 49 weeks and MRI image from the same patient.}
\end{figure}

In many research and clinical scenarios, researchers or clinicians often need to discern and analyze the disparities between two images captured at different time points or multimodal images acquired from the same patient. For instance, this could involve comparing different regions of interest in an ultrasound image to an MRI scan. Within our software (as illustrated in Figure \ref{fig:mutual_view}), we have seamlessly integrated this capability, empowering users to concurrently visualize and process two distinct images. This feature enhances the utility of MELAGE by enabling side-by-side examination and analysis of image pairs.

\subsubsection{Linking planes}
Coronal, sagittal and axial planes of a 3D image can be shown at the same time in MELAGE.
This ability helps the clinician to understand the exact localization of the ROI and is optimal for image segmentation as it allows visualization of the other 2 orthogonal planes when segmenting an image in one plane. 
In this way, an expert can work in one plane while simultaneously checking other planes to reduce the possible mistakes in segmentation.

\begin{figure}[H]
\centering
\subfigure[]{
\label{fig:linkage}\includegraphics[width=1\linewidth]{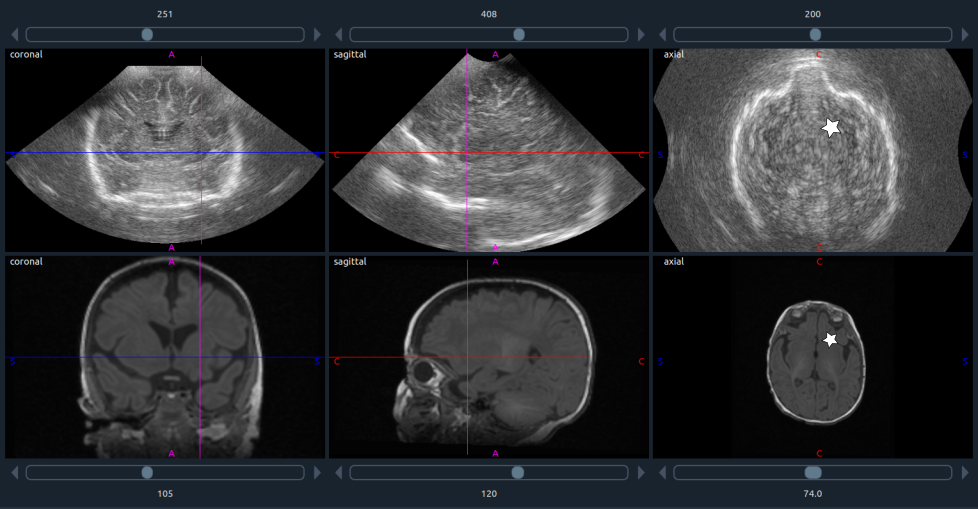}}
\caption{Linkage between planes; Star presents the location that has been touched by mouse in the axial plane.}
\end{figure}

Figure \ref{fig:linkage} shows the point located at coronal and sagittal planes while we touch a point in the axial plane for both MRI and Ultrasound images.

\subsubsection{Color Schemes}
\begin{figure}[H]
\centering
\subfigure[First color scheme]{
\label{fig:colrize0}\includegraphics[width=0.4\linewidth]{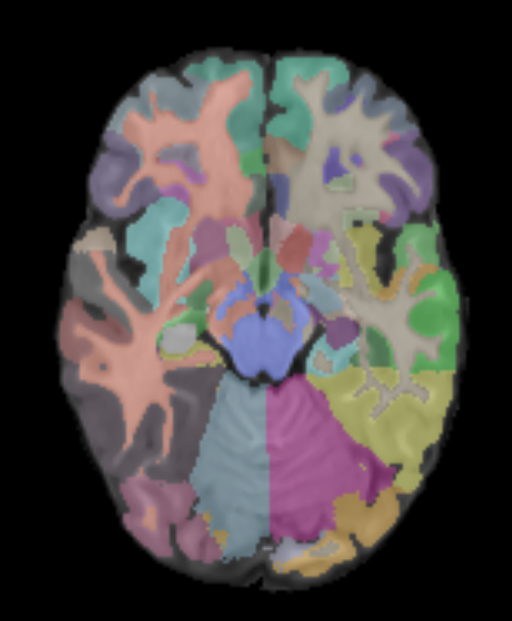}}
\subfigure[Second color scheme]{
\label{fig:colrize1}\includegraphics[width=0.4\linewidth]{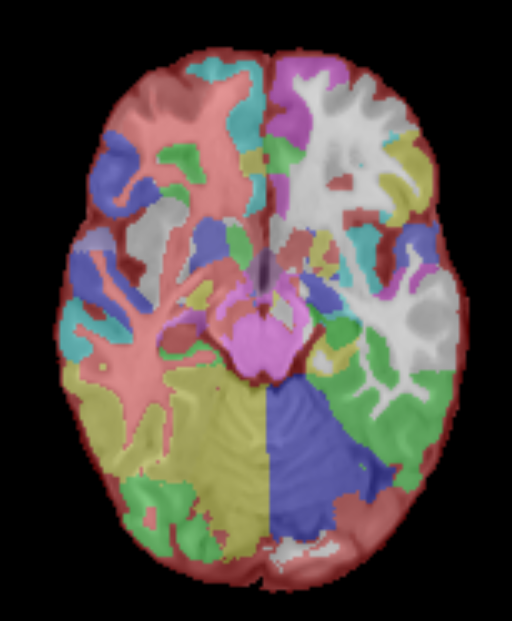}}
\caption{An example of a segmented image with different color schemes.}
\label{fig:colorschem}
\end{figure}

Various atlases utilize distinct color schemes for highlighting regions of interest. To offer users greater flexibility, we have incorporated the feature to select a color scheme that aligns with their preferences. Additionally, MELAGE allows users to introduce new custom color schemes, enabling dynamic visualization of the segmentation results. In Figure \ref{fig:colorschem}, you can observe a visual comparison between two distinct color schemes applied to the same segmentation, showcasing the versatility of our system.

\subsubsection{Instantaneous 3D visualization of the segmented ROIs}
\begin{figure}[H]
\centering
  \includegraphics[width=1.0\linewidth]{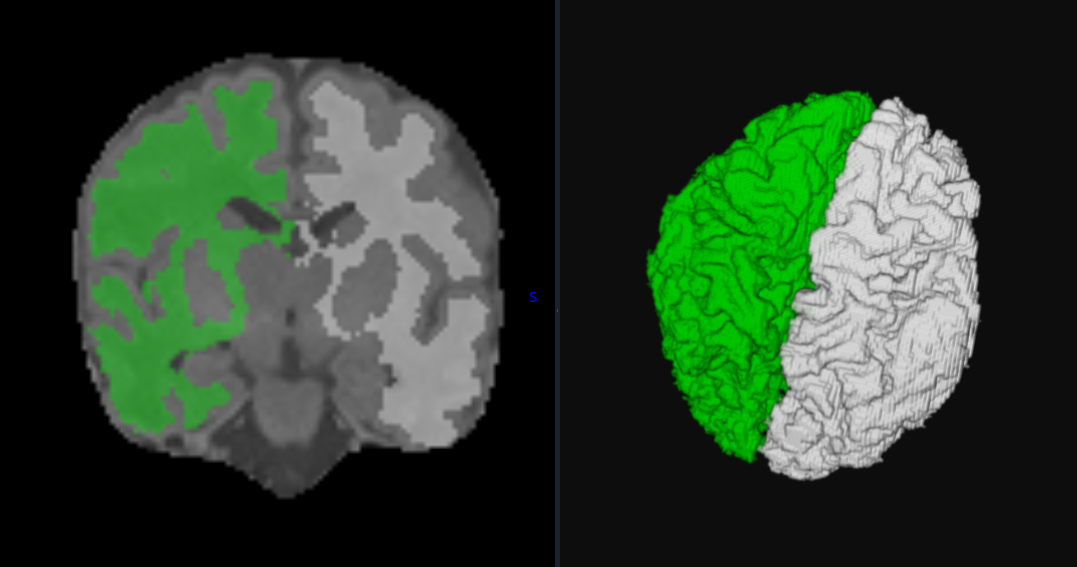}
  \caption{Instantaneous 3D visualization of the segmented ROIs during image segmentation, left: segmented left and right white matter. right: 3D visualization of the segmented image.}
  \label{fig_seg_inst}
\end{figure}

Enhanced visualization through 3D imaging provides a more comprehensive understanding of brain structures. As demonstrated in Figure \ref{fig_seg_inst}, we present an illustrative example of segmented white-matter in the coronal plane, accompanied by their corresponding 3D visualizations using consistent color coding. The real-time visualization of these segmented regions empowers users with direct control over their spatial representation, thereby enhancing segmentation precision and accuracy. This capability to interactively manipulate segmented structures in a 3D space not only aids in refining the segmentation process but also promotes a deeper insight into the intricate anatomy of the brain.

\subsubsection{Image Rotation, Zooming and Panning}
\begin{figure}[H]
\centering
\subfigure[Original image]{
\label{fig_rot_0}\includegraphics[width=0.4\linewidth]{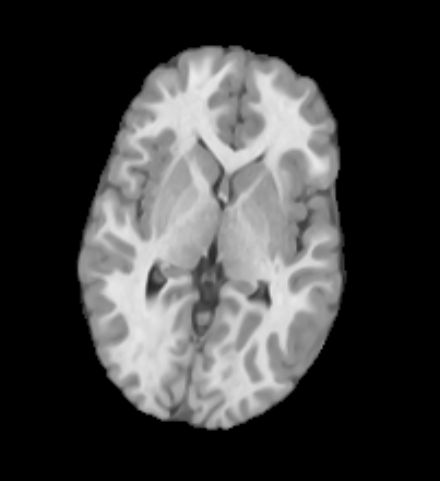}}
\subfigure[Rotated]{
\label{fig_rot_1}\includegraphics[width=0.4\linewidth]{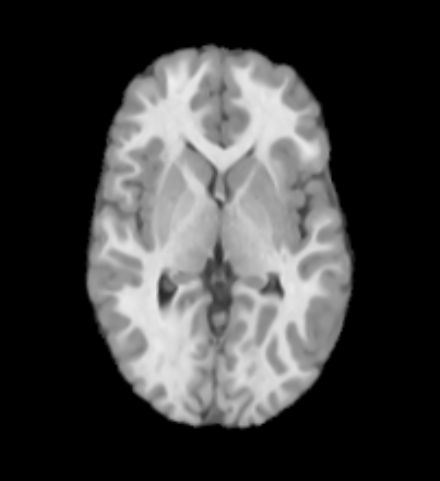}}\\
\subfigure[Original image]{
\label{fig_rot_2}\includegraphics[width=0.4\linewidth]{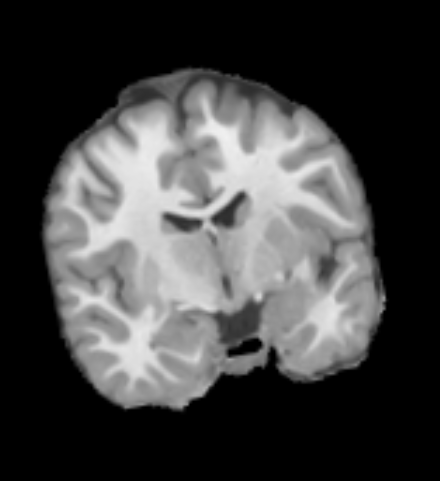}}
\subfigure[Rotated]{
\label{fig_rot_3}\includegraphics[width=0.4\linewidth]{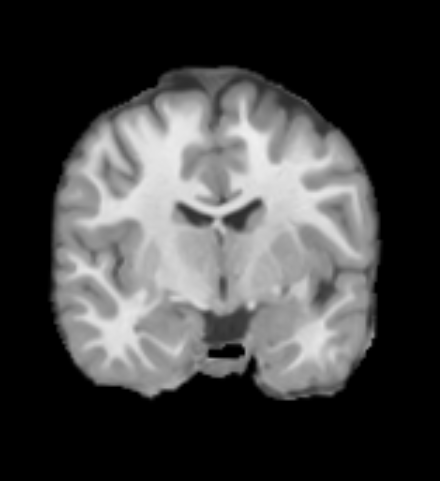}}\\
\subfigure[Original image]{
\label{fig_rot_4}\includegraphics[width=0.4\linewidth]{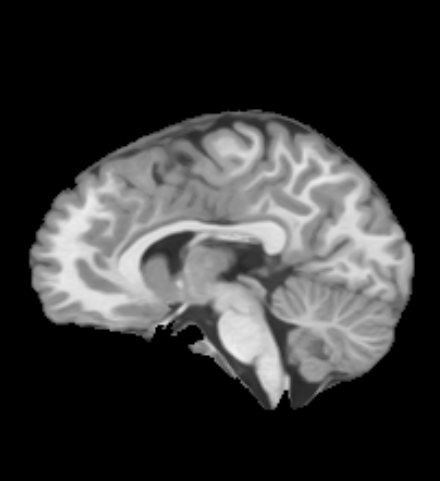}}
\subfigure[Rotated]{
\label{fig_rot_5}\includegraphics[width=0.4\linewidth]{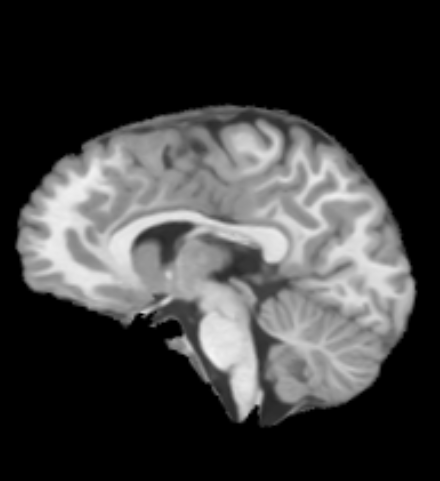}}
\caption{An example of 3D image rotation in Axial, Coronal and Sagittal planes using MELAGE platform.}
\label{fig_rot}
\end{figure}

Users have the capability to perform three-dimensional image rotations along axial, sagittal, and coronal planes. Figure \ref{fig_rot} provides an illustrative example showcasing the dynamic 3D image rotation across these planes. Subsequent to the desired rotation, the modified image can be effortlessly saved as a new file using the MELAGE platform. Additionally, MELAGE facilitates zooming and panning within the image, offering users the flexibility to segment images at their preferred zoom settings, thereby enhancing the precision and accuracy of the segmentation process.

\subsubsection{Supported image formats}
MELAGE showcases exceptional adaptability, providing seamless support for a multitude of image formats, including 3D NIfTI, DICOM, and NRRD. Furthermore, its versatility extends to the efficient management of 4D NIfTI images, making it an indispensable tool for comprehensive image analysis.
Remarkably, MELAGE boasts the capacity to process ultrasound images, encompassing data originating from both infants and fetal sources. This expansive coverage ensures robust and reliable image analysis across a diverse spectrum of medical imaging scenarios.
Additionally, MELAGE offers users the valuable functionality of accessing metadata contained within each file directly within the platform.

\subsubsection{Measurements}
MELAGE offers a comprehensive set of measurement tools for precise image analysis. Users can readily determine the distance between two points on an image and calculate the angle of the line connecting these reference points. During the segmentation process, MELAGE automatically computes both the area and perimeter of the segmented regions of interest (ROIs). These measurements are conveniently cataloged within the software, allowing users to save, and export the data to facilitate further analysis and documentation.

\begin{figure}[h!]\label{fig_seg_measure}
\centering
\subfigure[Linear measurements]{
\label{fig:seg_m_0}\includegraphics[width=0.4\linewidth]{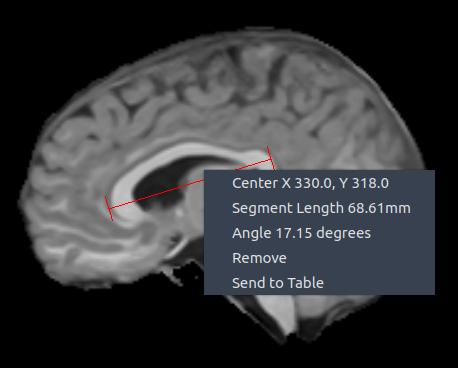}}
\subfigure[Segmented area measurements]{
\label{fig:seg_m_0}\includegraphics[width=0.4\linewidth]{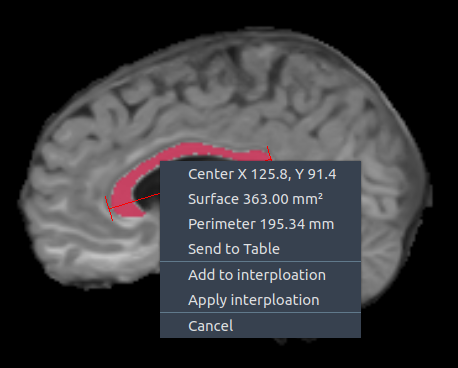}}
 \caption{ (a) Angle and distance measurements of an ROI or (b) area and perimeter measurement of the segmented ROI.}
\end{figure}

\subsubsection{Other features}
\begin{figure}[t!]
\centering
  \includegraphics[width=1.1\linewidth]{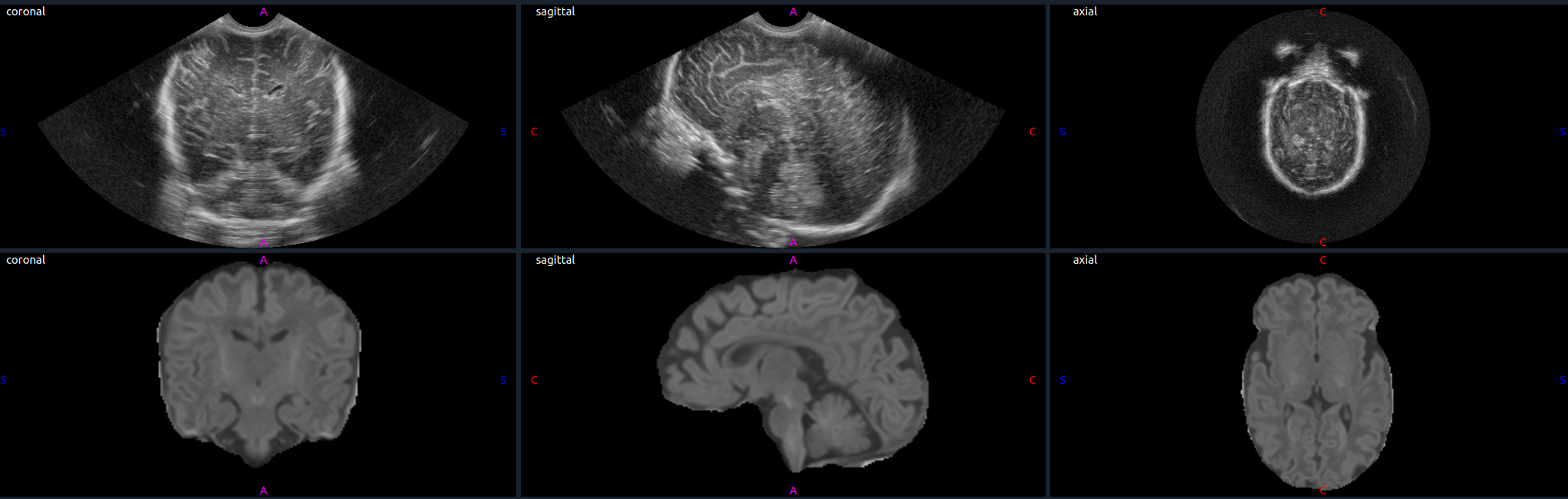}
  \caption{Showing and segmenting two image at the same time.}
  \label{fig_two_im0}
\end{figure}

The entire project, encompassing segmentation, measurements, and associated parameters, can be conveniently saved and later retrieved for continued work.
MELAGE offers the unique capability to simultaneously display and segment two distinct types of images, facilitating tasks such as image comparison and difference visualization (as demonstrated in Figure \ref{fig_two_im0}).
Another noteworthy feature of MELAGE is its proficiency in properly rendering ultrasound (US) planes. Given that US images often include sagittal or coronal planes, and 3D US scans can be acquired from similar orientations, MELAGE excels in accurately identifying and displaying images in their correct planes.
Furthermore, MELAGE empowers users with the ability to visualize image histograms and manipulate image coordinate systems to select the most suitable ones. Notably, it integrates the N4 bias field correction from the SimpleITK package \cite{yaniv2018simpleitk} to enhance image quality.
MELAGE also offers advanced masking functionalities, allowing users to mask specific regions of an image in relation to segmentation. This includes operations like summation and subtraction of segmentation colors, enabling comprehensive image manipulation.

\subsection{Advanced features}
The advanced features are presented in software called MELAGE.

\subsubsection{Interactive image segmentation}
One of the critical aspects of manual segmentation or correction of automated segmentation is the provision of an interactive tool for image segmentation. This tool should effectively assist users in selecting the precise locations for segmentation.

\begin{figure}
  \includegraphics[width=\linewidth]{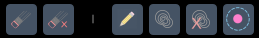}
  \caption{Segmentation toolbar of MELAGE.}
  \label{fig:segtool}
\end{figure}

In response to the specific requirements of medical professionals from the "Perinatal brain damage" research group at the Biomedical Research and Innovation Institute of Cádiz ( \href{https://inibica.es/co20-dano-cerebral-perinatal/}{INiBICA}) and the Puerta del Mar University Hospital, Cádiz, Spain, we have designed an easily navigable menu bar within MELAGE. This menu bar incorporates tools tailored to the needs of medical doctors. We have implemented six distinct tools: three for image segmentation, one for removing unwanted segmentation, and two for applying segmentation to subsequent slices (refer to Figure \ref{fig:segtool}).

\begin{figure}
\centering
  \includegraphics[width=0.5\linewidth]{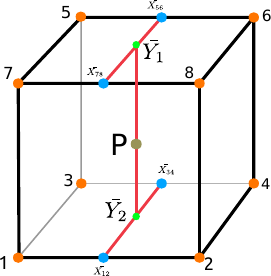}
  \caption{Trilinear interpolation.}
  \label{fig:tril}
\end{figure}

MELAGE incorporates several advanced capabilities, each designed to enhance the process of image segmentation and analysis:

\begin{itemize}
    \item \textbf{Polygon Construction and Filling:}  One of the fundamental features in MELAGE is the ability to construct polygons using a set of user-selected points, denoted as $\mathbf{X}$. These interconnected points collectively form a polygon, referred to as $P$. Once the polygon is defined, MELAGE offers a straightforward method for filling it with a user-defined color. The process begins by creating a bounding box, represented as $B$, around the selected points $\mathbf{X}$. Subsequently, all the points located within this polygon are marked and filled with the chosen color. This intuitive feature streamlines the segmentation process, enabling users to highlight and annotate precisely regions of interest.
    \item \textbf{Interpolation Between Slices:} MELAGE offers a valuable capability for scenarios where two or more slices are segmented from the same image plane. To ensure a seamless transition between these segmented slices, MELAGE calculates the exact Euclidean distance for each pixel within every segmented slice. The Euclidean distance, represented as $y_{i}$, is computed as follows:
    \begin{equation}
    y_{i} = \sqrt{ \sum_{i=1}^{N} (x_{i} - b_{i} )^{2} }
    \end{equation}
    Here, $x_{i}$ denotes the coordinates of each pixel, and $b_{i}$ signifies a baseline point. With precise Euclidean distances determined, MELAGE employs the advanced trilinear interpolation method \cite{bourke1999interpolation} to seamlessly bridge the gap between two segmented slices. This interpolation technique harnesses neighboring points, as illustrated in Figure \ref{fig:tril}, to estimate the values of points situated between the two segmented slices. By doing so, We ensure a continuous and smooth representation of the segmented points. This capability is particularly valuable for maintaining visual consistency and coherence when working with multiple segmented slices from the same image plane.
\end{itemize}
These innovative features collectively position MELAGE as a versatile and powerful tool for image segmentation and analysis. Whether users are creating and filling polygons or seamlessly interpolating between image slices, MELAGE empowers them with a suite of tools to enhance precision and efficiency in their image processing endeavors.

In the realm of medical image segmentation, there are instances where the task at hand necessitates the precise delineation of regions within an image characterized by uniform intensity levels. To address this specific requirement, we introduces a dedicated tool inside MELAGE to streamline this process within regions of interest (ROI). This tool operates by defining a circular ROI centered at a user-specified point, with the radius denoted as $R$. Importantly, users have the flexibility to adjust this radius ($R$) according to their specific needs and preferences.
The pixel selection process within this circular ROI is governed by a set of predefined conditions, ensuring that only pixels meeting these criteria are included in the segmentation. Specifically, the following conditions guide the selection:
\begin{equation}
    M_{c} - Std_{l} <S< M_{c} + Std_{l}
\end{equation}
where $M_{c}$ is the intensity value of the voxel located at the center of the image. This central reference point serves as the baseline for assessing intensity uniformity within the defined circular ROI. $Std_{l}$ represents the standard deviation of intensity values among voxels situated within the circular region. This standard deviation metric plays a pivotal role in evaluating the extent of intensity variation observed within the specified ROI. By adhering to these defined parameters, MELAGE effectively identifies and selects pixels residing within the circular ROI that adhere to the specified intensity criteria. This segmentation approach facilitates the accurate isolation of regions characterized by consistent intensity levels, affording users a precise and adaptable means of defining areas of interest within medical images.

\begin{figure}[h!]
\centering
\subfigure[]{
\label{fig:seg_a}\includegraphics[width=0.3\linewidth]{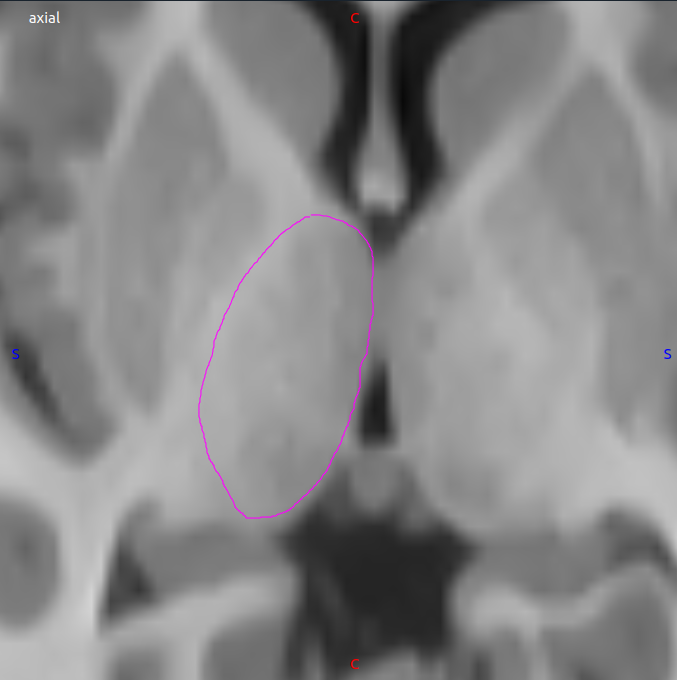}}
\subfigure[]{
\label{fig:seg_b}\includegraphics[width=0.3\linewidth]{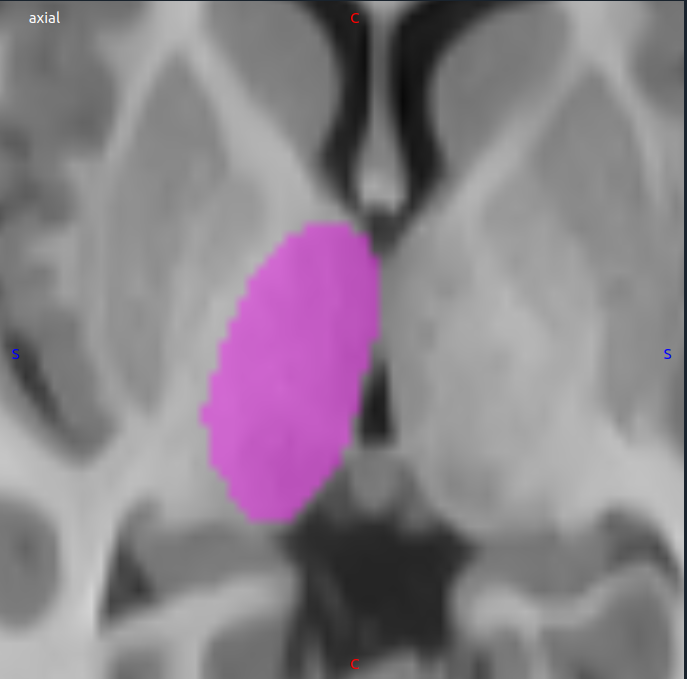}}

\caption{Segmentation of thalamus in T1 image using 1) Contour segmentation }
\end{figure}

\begin{figure}[H]
\centering
\subfigure[]{
\label{fig:seg_a}\includegraphics[width=0.3\linewidth]{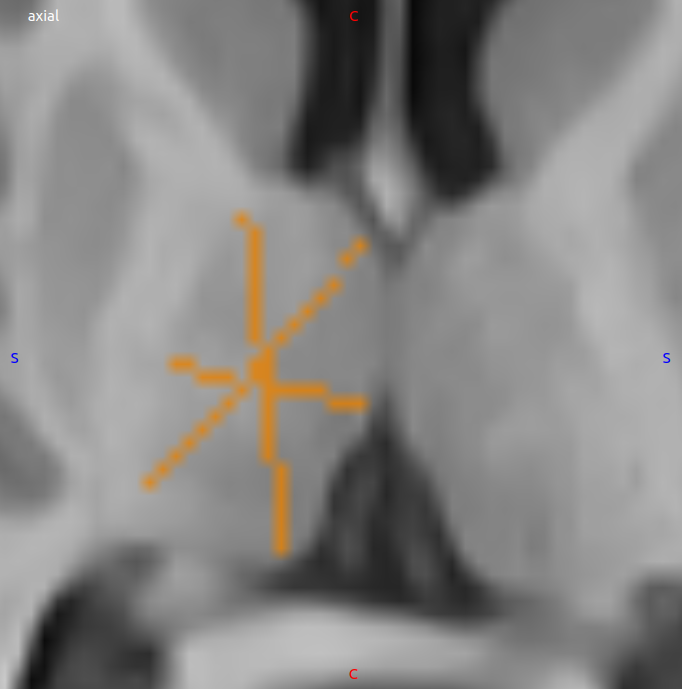}}
\subfigure[]{
\label{fig:seg_b}\includegraphics[width=0.3\linewidth]{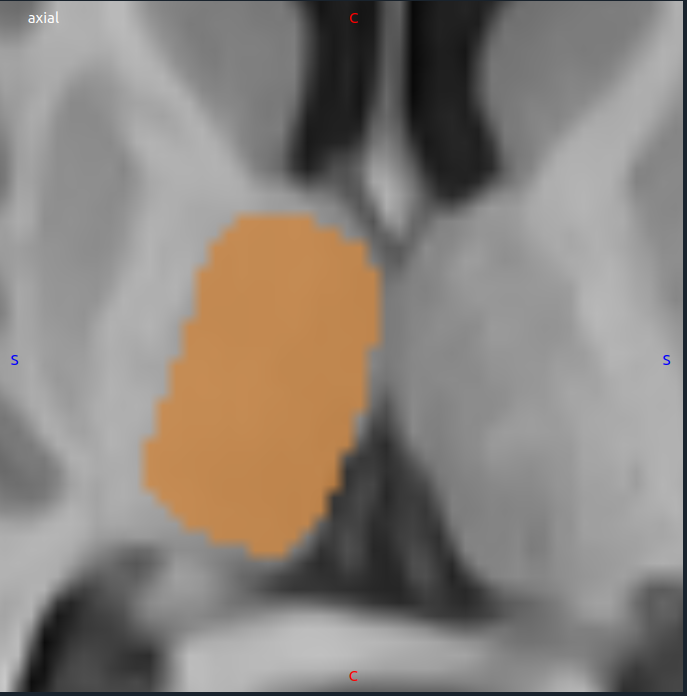}}

\caption{Segmentation of thalamus in a T1 image using 1) Contour segmentation }
\end{figure}

\subsubsection{Editing segmented ROIs}
\begin{figure}[h!]
\centering
\includegraphics[width=0.9\linewidth]{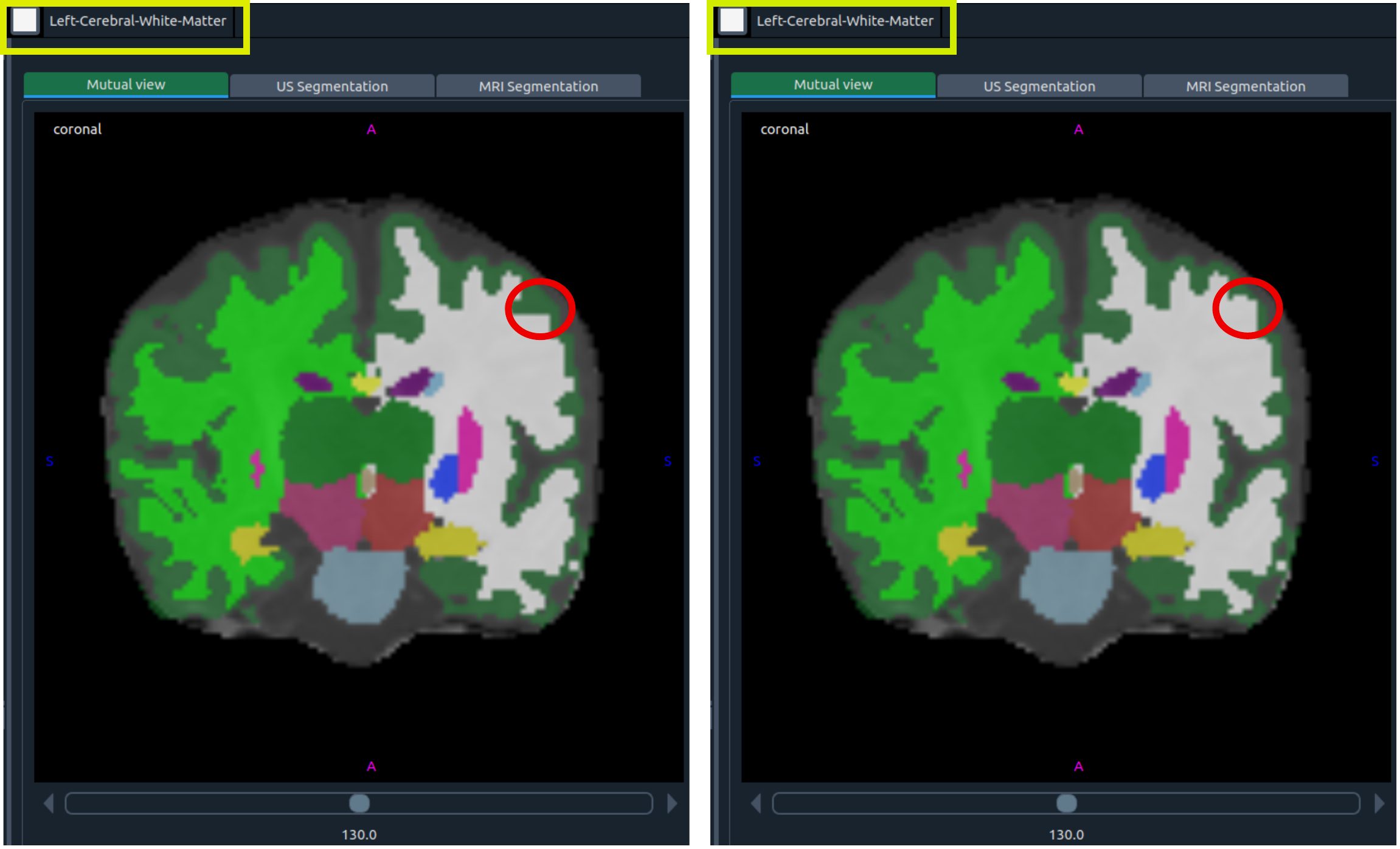}
\caption{Example of editing segmentation.}
 \label{fig:segedit}
\end{figure}

MELAGE offers users the ability to not only segment regions of interest but also edit and refine segmented ROIs as needed. Figure \ref{fig:segedit} serves as an illustrative example of the segmentation editing process. Prior to segmentation, users are required to select the appropriate label corresponding to the target region. In this particular instance, the objective is to modify the segmentation of the left cerebral white matter. Subsequently, the corresponding color, white, is designated (located at the top left of Figure \ref{fig:segedit}). The subsequent step involves overwriting any extraneous gray matter with the designated white matter color (right portion of Figure \ref{fig:segedit}). Notably, the edited region is prominently highlighted with a distinctive red circle for clarity.

\subsubsection{Automatic brain extraction}
Within the framework of MELAGE, a dedicated tool has been developed to automate the segmentation of neonatal brain structures from MRI and 3D Ultrasound images. The efficacy of this tool has undergone rigorous validation through an extensive dataset comprising images acquired from HUPM and several other independent centers. The underlying network architecture employed is U-Net, meticulously trained on a substantial dataset, encompassing over 500 MRI images sourced from HUPM and an additional 150 ultrasound (US) images from the HUPM dataset.
We have incorporated the code described in \cite{hoopes2022synthstrip} to perform MRI brain extraction. Users have the flexibility to adjust the threshold and review and refine the segmentation outcomes using MELAGE.

\begin{figure}[h!]
\centering
\includegraphics[width=0.9\linewidth]{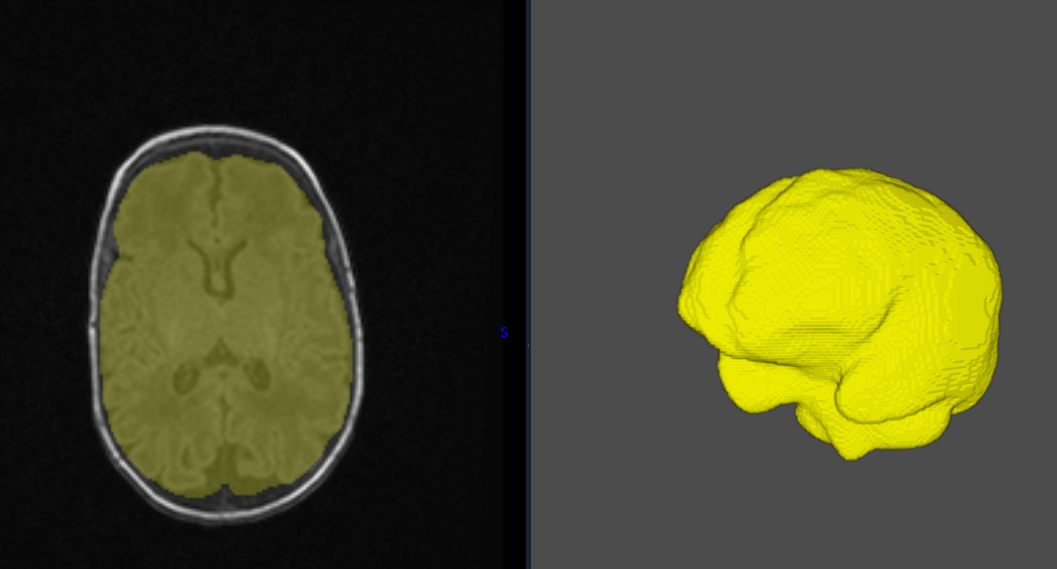}
\caption{Example of automatic brain segmentation.}
 \label{fig:brainext}
\end{figure}

Figure [\ref{fig:brainext}] illustrates a representative result obtained using the MELAGE software, showcasing the robustness and accuracy of the segmentation process. Further comprehensive details regarding the methodology and findings are slated for imminent publication, promising to contribute significantly to the field of neuroimaging.

\section{Future perspective} \label{future}
It is easy to integrate MELAGE with deep learning algorithms and other artificial intelligence methods.
We will add artificial intelligence methods for automatic analysis of 3D MRI and Ultrasound images.
The automatic analysis of Diffusion Weighted Imaging (DWI) will be another capability of MELAGE in future.
\section{Conclusion}\label{conclusion}
In summary, MELAGE represents a pioneering advancement in the field of neuroimaging, offering a pure Python-based software solution designed for the visualization, processing, and analysis of medical images. Originally conceived for the specific purpose of processing 3D ultrasound and MRI brain images in neonates, MELAGE exhibits remarkable versatility, extending its capabilities to encompass adult human brain imaging.

Central to MELAGE's functionality is its semi-automatic brain extraction tool, fortified by an embedded deep learning module adept at extracting intricate brain structures from MRI and 3D Ultrasound images. Beyond this, MELAGE offers a comprehensive suite of features, including dynamic 3D visualization, precise measurements, and interactive image segmentation.

What sets MELAGE apart is its innate adaptability and ease of integration with deep learning algorithms and other artificial intelligence methods. This adaptability not only positions MELAGE at the forefront of neuroimaging innovation but also makes it a potent tool for interdisciplinary research and clinical applications.

Furthermore, MELAGE's utility extends beyond its original scope, as it can effectively process various other types of medical image data, underscoring its potential as a versatile resource for the broader medical imaging community. As we look to the future, MELAGE stands as a beacon of innovation, poised to catalyze new discoveries, streamline analysis, and foster collaborative advancements in the realm of medical imaging and beyond.

\section*{Acknowledgement}
This study was funded  by the  Cadiz integrated territorial initiative for biomedical research, European Regional Development Fund (ERDF) 2014–2020. Andalusian Ministry of Health and Families, Spain. Registration number: ITI-0019-2019.
This study has been funded by Instituto de Salud Carlos III (ISCIII) through the project "DTS22/00142" and co-funded by the European Union.
We extend our heartfelt gratitude to our collaborators in this project, including:Lionel Cervera Gontard, Pedro Luis Galindo Riaño and Joaquin Pizarro Junquera. We also extend our appreciation to numerous technicians, master's, and Ph.D. students who helped us in the development of MELAGE: Manuel Lubián Gutiérrez, Ph.D. Student, Emiliano Trimarco, Ph.D. Student, Roa’a Khaled, Ph.D. Student, Monica Crotti, Ph.D. Student, Yolanda Marín Almagro, Ph.D. Student, Macarena Román, Technician. Your invaluable contributions have significantly enriched this project.

\section*{Code Availability}
MELAGE is a free software and the open source code of the main body is available through Github\footnote{\href{https://github.com/BahramJafrasteh/MELAGE}{https://github.com/BahramJafrasteh/MELAGE}}. For more information please see \href{https://melage.uca.es/}{https://melage.uca.es/}.

\section*{Registration}
MELAGE has been registered \footnote{\href{https://www.safecreative.org/work/2211222681375-melage}{https://www.safecreative.org/work/2211222681375-melage}} under identifier 2211222681375.

\bibliography{mybibfile}
\bibliographystyle{elsarticle-num}

\end{document}